\documentclass[letterpaper,twocolumn,fleqn]{article} 

\usepackage{ist}
\usepackage{amsmath}
\usepackage[pagebackref=true,breaklinks=true,letterpaper=true,colorlinks,bookmarks=false]{hyperref}
\usepackage{subcaption}

\title{Attribution of Gradient Based Adversarial Attacks for Reverse Engineering of Deceptions}

\author{Michael Goebel\textsuperscript{1}, 
Jason Bunk\textsuperscript{2}, 
Srinjoy Chattopadhyay\textsuperscript{2}, 
Lakshmanan Nataraj\textsuperscript{2}, 
Shivkumar Chandrasekaran\textsuperscript{1,2}, 
B. S. Manjunath\textsuperscript{1,2}
\\
\textsuperscript{1}University of California, Santa Barbara \\
\textsuperscript{2}Mayachitra Inc; Santa Barbara, CA}

\date{} % date has an empty field.

\begin{document} 

\maketitle 

\thispagestyle{empty} % prevents the first page to be numbered

%%%%%%%%%%%%%%%%%%%%%%%%%%%%%%%%%%
% Abstract
%%%%%%%%%%%%%%%%%%%%%%%%%%%%%%%%%%

\begin{abstract}
Machine Learning (ML) algorithms are susceptible to adversarial attacks and deception both during training and deployment.
Automatic reverse engineering of the toolchains behind these adversarial machine learning attacks will aid in recovering the tools and processes used in these attacks. 
In this paper, we present two techniques that support automated identification and attribution of adversarial ML attack toolchains using Co-occurrence Pixel statistics and Laplacian Residuals. 
Our experiments show that the proposed techniques can identify parameters used to generate adversarial samples. To the best of our knowledge, this is the first approach to attribute gradient based adversarial attacks and estimate their parameters. 
Source code and data is available at: https://github.com/michael-goebel/ei\_red.
\end{abstract}

%%%%%%%%%%%%%%%%%%%%%%%%%%%%%%%%%%%%
% Overall Document Guidelines: Head
%%%%%%%%%%%%%%%%%%%%%%%%%%%%%%%%%%%%

\section{Introduction}
\label{sec:intro}

Convolutional neural networks (CNNs) are increasingly being used in critical applications, such as self-driving cars and face authentication. Recent works have shown that gradient based attacks can reduce accuracy of visual recognition networks to less than 1\%, while minimally perturbing an image. The adversary uses gradient descent through the network to maximize the output at an incorrect label, while minimizing the perturbation to the image. Various attack methods have been produced using this common framework, including Fast Gradient Sign Method (FGSM) \cite{goodfellow2014explaining} and Projected Gradient Descent (PGD) \cite{madry2017towards}. Works have also been proposed to detect such adversarial samples, but none have been published which can estimate the adversarial setup from image samples. Knowing such parameters would allow for more accurate adversarial retraining against such attacks as well as aid in recovering the tools and processes used in these attacks~\cite{darpa-red}.

\begin{figure}[h]
    \centering
    \includegraphics[width=\linewidth]{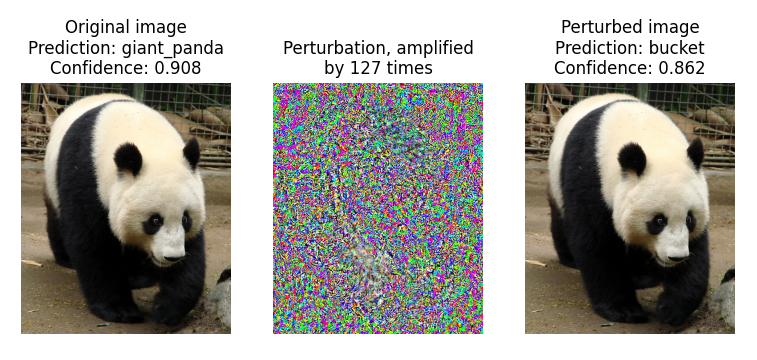}
    \caption{A sample PGD attack against ResNet. Small perturbations against a network with known weights can lead to significant differences in prediction outputs. Scores indicated here are confidence scores from 0-1, where the sum of all scores is equal to 1.}
    \label{fig:sample_attack}
\end{figure}

Gradient-descent based adversarial attacks use the gradients of deep neural networks (DNNs) to imperceptibly alter their inputs so as to change the output dramatically. Within this family, there are various strains of algorithms, each with several parameters. In this work, we propose to detect such adversarial attack toolchains and their parameters. Our objectives are two-fold:

\begin{enumerate}

\item To attribute an adversarially attacked  image to a particular attack toolchain/family, 

\item Once an attack has been identified, determine the parameters of the attack so as to facilitate the reverse engineering of these adversarial deceptions. 

\end{enumerate}

We will now briefly describe some of the attacks considered for detection and attribution. A deep neural network (DNN) is represented as a function $f: X \rightarrow Y$, where $X$ denotes the input space of data and $Y$ denotes the output space of the classification categories. The training set comprises known pairs $(x_t,y_t)$, where $x_t \in X$ and $y_t \in Y$,and $f()$ is obtained by minimizing a loss function $J(f(x_t),y_t)$. We will consider the following attacks:

\begin{enumerate}
\item Fast Gradient Sign Method (FGSM): This attack perturbs a clean image x by taking a fixed step  in the direction of the gradient of $J(f(x_t),y_t)$ with respect to $x_t$.
\item Projected Gradient Descent (PGD): This attack is an improvement over FGSM, where the adversarial samples $x'$ are generated by multiple iterations and intermediate results are clipped so as to keep them within the $\epsilon$-neighborhood of $x:{x'}_i = {x'}_{i-1} - {clip}_{\epsilon}(\alpha \cdot sign({\nabla}_x J(f({x'}_{i-1},y)) )$.

\end{enumerate}

These two attacks are examples of $l_{\infty}$ attacks, where $\epsilon$ represents the maximum allowable perturbation to any pixel in x. The software repositories of these attacks can be obtained from  the following: Advertorch~\cite{ding2019advertorch}, Adversarial Robustness Toolbox~\cite{nicolae2018adversarial}, Foolbox~\cite{rauber2017foolbox}, CleverHans~\cite{papernot2016technical}. A PGD example from the Advertorch toolbox is given in Figure \ref{fig:sample_attack}.

\section{Related Works}

Many works have taken the approach of creating more robust networks, for which small changes in input will not significantly change the output classification \cite{bastani2016measuring,gu2014towards,huang2015learning,jin2015robust,papernot2016distillation,rozsa2016adversarial,shaham2015understanding,zheng2016improving}. Generally, these methods cause a significant decrease in accuracy, for both tampered and untampered images \cite{carlini2017adversarial}. While these networks are necessary when class estimation is required for all samples, others methods may be more favorable when this requirement is relaxed.

Detection has become another popular approach to circumventing these attacks \cite{bhagoji2017dimensionality,feinman2017detecting,gong2017adversarial,grosse2017statistical,metzen2017detecting,hendrycks2016early}. Such methods allow for the classification networks to remain as is, while filtering out adversarial examples before they reach the target network. The methods presented in this paper move a step beyond simple detection, with the addition of attack classification and parameter estimation.

\begin{figure}[t]
    \centering
    \includegraphics[width=\linewidth]{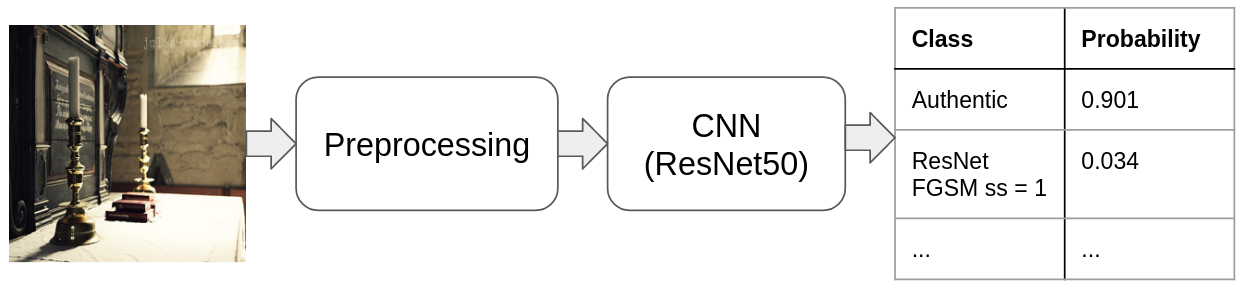}
    \caption{High level model diagram for detection. All models fit into this framework, with different preprocessing methods.}
    \label{fig:model}
\end{figure}

\section{Method}

\subsection{Model}

To enhance the artifacts created by adversarial attacks, we consider two preprocessing methods common to image forensic, before training a neural network.
A visual summary of our detector is given in Figure \ref{fig:model}.
As a baseline, we compare these two methods against a method with no preprocessing. The first is a Laplacian high-pass filter. Similar filters have been used for both image resampling detection \cite{kirchner2008fast}, and general image manipulation detection \cite{bayar2016deep}. In our tests, the following 3x3 filter was applied to each of the RGB channels:

\begin{equation}
    h(x,y) = 
    \begin{bmatrix}
    1 & 1 & 1 \\
    1 & -8 & 1 \\
    1 & 1 & 1
    \end{bmatrix}
\end{equation}

\begin{figure}[t]
    \centering
    \includegraphics[width=0.8\linewidth]{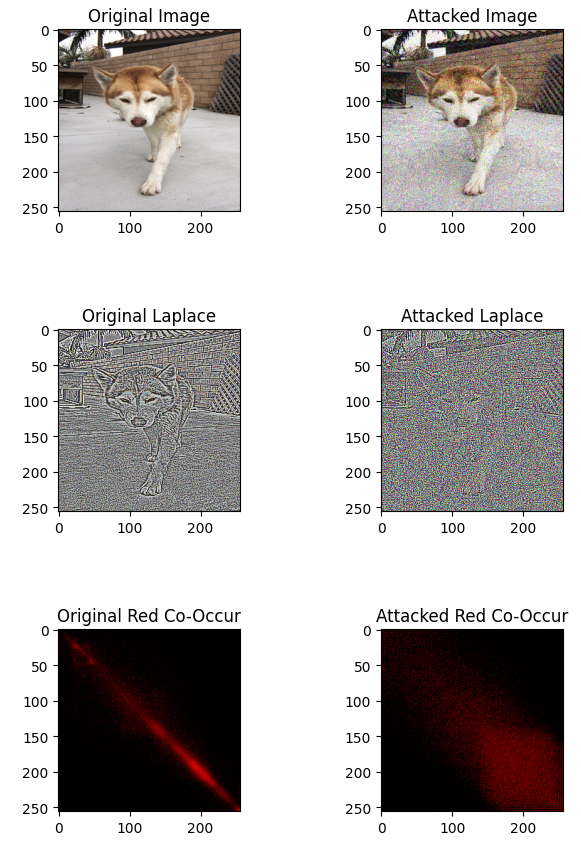}
    \caption{An untampered image, and corresponding PGD attacked image, with a large step size and number of steps to amplify the difference. The adversarial noise added appears across the whole image. The difference in the co-occurrence matrices is notable in the significant increase in spread about the diagonal.}
    \label{fig:example_inputs}
\end{figure}

The second preprocessing method investigated is the co-occurrence matrix. Such matrices have been used extensively in detection of steganography \cite{sullivan2005steganalysis,sullivan2006steganalysis} as well as in detection of GAN images \cite{nataraj2019detecting}. For this method, two dimensional histograms of adjacent pixel pairs are constructed for each of the color channels. Below we show the equation for horizontal pairs, where X is a 2D array representing a single color channel. A sample image passed through each mode of processing is shown in Figure \ref{fig:example_inputs}.

\begin{equation}
    C_{i,j} = \sum_{m,n} [X_{m,n} = i][X_{m,n+1} = j]
\end{equation}

This can be applied to $X^T$ for vertical pairs as well, and on all three channels. These 6 co-occurrence matrices are then stacked into a final input tensor of size $256 \times 256 \times 6$. This tensor is passed to a CNN classifier as a multi-channel image.

\subsection{Detection, Attribution, and Estimation}

\begin{figure*}[t]
    \centering
    \includegraphics[width=0.9\linewidth]{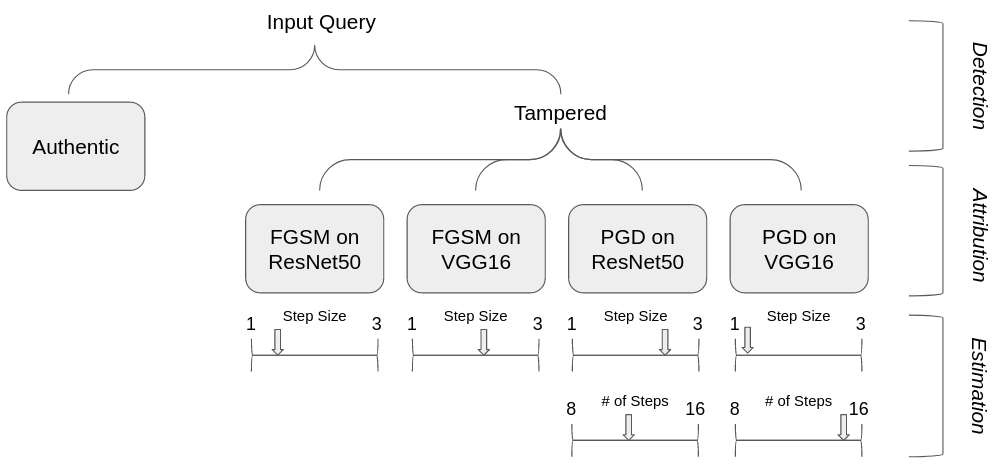}
    \caption{Levels of information provided to the user by our method. Using a single network, we demonstrate results for detection, attribution, and parameter estimation.}
    \label{fig:flowchart}
\end{figure*}

For our final output, we would like to tell a user whether or not a query is tampered, what attack method was used, and the parameters for that method. This high level idea described visually in Figure \ref{fig:flowchart}.
To accomplish this, we train a multiclass network, with each attack and parameter combination as a different label. To form the aggregated sets, such as real vs tampered, we sum the model outputs associated with each set. The set with the largest output is selected as the estimated class. 

If the image is predicted to be tampered, we then compute our parameter estimates using the model outputs for the predicted meta-class. A weighted sum is used, with the model outputs as the weights, and the associated class parameters as the values.

\begin{equation}
    P_{est} = \frac{\sum_{i \in S} P_i \times y_i}{\sum_{i \in S} y_i}
    \label{eq:weighted_sum}
\end{equation}

\section{Experiments}

\subsection{Dataset}

A full list of the attacks investigated is given in table \ref{tab:attacks_summary}. These attacks are repeated on VGG16 and ResNet50, and each is classified separately. Only the ends of the parameter spectrum are used for training. Parameters which are in-between these ends of the spectrum are seen only at test time. A total of 12 different tampered classes are used at training time, with one additional class for untampered, for a total of 13. 

The dataset is constructed from a random selection of ImageNet samples, all resized to $256 \times 256$. Attacks are run as targeted, with the new label being randomly selected from the 999 labels which are different than the associated ground truth. The attacks are then run to maximize the network output for the target class.

\begin{table}[]
 \centering
    \begin{tabular}{|c|c|c|c|}
        \hline
        Attack & Parameters & Training & Testing \\
        \hline
        FGSM & ss = 1 & X & X \\
        FGSM & ss = 2 &  & X \\
        FGSM & ss = 3 & X & X \\
        PGD & ss = 1, ns = 8 & X & X \\
        PGD & ss = 1, ns = 12 &  & X \\
        PGD & ss = 1, ns = 16 & X & X \\
        PGD & ss = 2, ns = 8 &  & X \\
        PGD & ss = 2, ns = 12 &  & X \\
        PGD & ss = 2, ns = 16 &  & X \\
        PGD & ss = 3, ns = 8 & X & X \\
        PGD & ss = 3, ns = 12 &  & X \\
        PGD & ss = 3, ns = 16 & X & X \\
        \hline
    \end{tabular}
    \vspace{6pt}
    \caption{Breakdown of the attacks used for training and testing. All attacks are repeated against pretrained VGG16 and ResNet50. "ss" denotes "stride size", assuming an image is in the range [0,255], and "ns" denotes "number of steps".}
    \label{tab:attacks_summary}
\end{table}

\subsection{Model Training}

A ResNet50 pretrained on ImageNet was used as our initial network, with the input and output layers modified to accommodate the different input and output sizes for this task. The model was trained over 20 epochs, using a batch size of 32, Adam optimizer, and cross-entropy loss. After each of the epochs, the model was evaluated on the validation set. The weights corresponding to the lowest validation loss were saved, and used for the remainder of the tests.

\subsection{Results}

\begin{figure}[!h]
    \centering
    \includegraphics[width=0.8\linewidth]{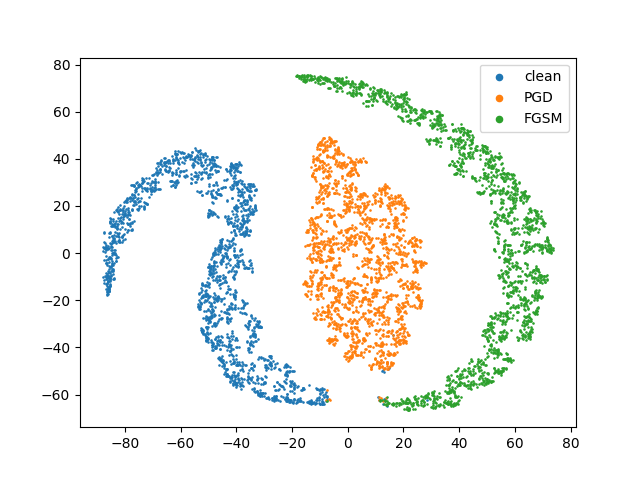}
    \caption{t-SNE results from the co-occurrence model for one of the classification tasks. Features are taken from the penultimate layer of the detection network, and run though t-SNE dimensionality reduction. Clear clusters are seen dividing each class.}
    \label{fig:tsne}
\end{figure}

Table \ref{tab:classification} shows our results for several different separations of the meta-classes. The co-occurrence and direct methods performed better on average than the Laplace method across the different tasks. Considering all 3 detectors, our methods achieved at least 90\% accuracy for each task. Figure \ref{fig:tsne} shows a t-SNE clustering of the deep features taken from one of these classification networks.

Table \ref{tab:estimation} shows the results of each method on different estimation tasks. Notably, the Laplace and co-occurrence methods out-perform the baseline direct method in several of the estimation tasks.

\begin{table*}[]
    \centering
    \scriptsize
    
\begin{tabular}{|c|c|c|c|c|c|c|c|c|c|c|c|c|c|c|}
\hline
original\_resized & 0.864 & 0.012 & 0.001 & 0.001 & 0.000 & 0.000 & 0.000 & 0.110 & 0.007 & 0.005 & 0.000 & 0.000 & 0.000\\
resnet50\_FGSM\_ss\_1 & 0.026 & 0.762 & 0.004 & 0.001 & 0.000 & 0.000 & 0.000 & 0.177 & 0.021 & 0.008 & 0.000 & 0.000 & 0.000\\
resnet50\_FGSM\_ss\_3 & 0.000 & 0.002 & 0.977 & 0.015 & 0.000 & 0.000 & 0.000 & 0.000 & 0.004 & 0.003 & 0.000 & 0.000 & 0.000\\
resnet50\_PGD\_ns\_8\_ss\_1 & 0.000 & 0.000 & 0.042 & 0.953 & 0.000 & 0.003 & 0.000 & 0.000 & 0.000 & 0.001 & 0.000 & 0.000 & 0.000\\
resnet50\_PGD\_ns\_8\_ss\_3 & 0.000 & 0.000 & 0.000 & 0.000 & 0.999 & 0.000 & 0.000 & 0.000 & 0.000 & 0.000 & 0.000 & 0.000 & 0.000\\
resnet50\_PGD\_ns\_16\_ss\_1 & 0.000 & 0.000 & 0.000 & 0.008 & 0.003 & 0.988 & 0.000 & 0.000 & 0.000 & 0.000 & 0.000 & 0.001 & 0.000\\
resnet50\_PGD\_ns\_16\_ss\_3 & 0.000 & 0.000 & 0.000 & 0.000 & 0.001 & 0.000 & 0.999 & 0.000 & 0.000 & 0.000 & 0.000 & 0.000 & 0.000\\
vgg16\_FGSM\_ss\_1 & 0.244 & 0.018 & 0.001 & 0.001 & 0.000 & 0.000 & 0.000 & 0.704 & 0.023 & 0.009 & 0.000 & 0.000 & 0.000\\
vgg16\_FGSM\_ss\_3 & 0.007 & 0.007 & 0.004 & 0.000 & 0.000 & 0.000 & 0.000 & 0.047 & 0.848 & 0.087 & 0.000 & 0.000 & 0.000\\
vgg16\_PGD\_ns\_8\_ss\_1 & 0.000 & 0.001 & 0.002 & 0.001 & 0.000 & 0.000 & 0.000 & 0.002 & 0.003 & 0.989 & 0.000 & 0.002 & 0.000\\
vgg16\_PGD\_ns\_8\_ss\_3 & 0.000 & 0.000 & 0.000 & 0.000 & 0.000 & 0.001 & 0.000 & 0.000 & 0.000 & 0.000 & 0.989 & 0.009 & 0.001\\
vgg16\_PGD\_ns\_16\_ss\_1 & 0.000 & 0.000 & 0.000 & 0.000 & 0.000 & 0.000 & 0.000 & 0.000 & 0.000 & 0.003 & 0.003 & 0.993 & 0.000\\
vgg16\_PGD\_ns\_16\_ss\_3 & 0.000 & 0.000 & 0.000 & 0.000 & 0.000 & 0.000 & 0.000 & 0.000 & 0.000 & 0.000 & 0.000 & 0.000 & 1.000 \\
\hline
\end{tabular}

    \vspace{6pt}
    \caption{Confusion matrix for direct method on test dataset. Column labels are in the same order as row labels. Rows indicate ground truth, columns indicate predicted.}
    \label{tab:conf_dir}
\end{table*}

\begin{table*}[]
    \centering
    \scriptsize
    
\begin{tabular}{|c|c|c|c|c|c|c|c|c|c|c|c|c|c|c|}
\hline
original\_resized & 0.919 & 0.064 & 0.003 & 0.004 & 0.000 & 0.001 & 0.000 & 0.008 & 0.000 & 0.000 & 0.000 & 0.000 & 0.000\\
resnet50\_FGSM\_ss\_1 & 0.035 & 0.943 & 0.007 & 0.005 & 0.000 & 0.001 & 0.000 & 0.008 & 0.001 & 0.000 & 0.000 & 0.000 & 0.000\\
resnet50\_FGSM\_ss\_3 & 0.000 & 0.007 & 0.981 & 0.009 & 0.000 & 0.000 & 0.000 & 0.000 & 0.002 & 0.000 & 0.000 & 0.000 & 0.000\\
resnet50\_PGD\_ns\_8\_ss\_1 & 0.000 & 0.000 & 0.000 & 0.999 & 0.000 & 0.000 & 0.000 & 0.000 & 0.000 & 0.000 & 0.000 & 0.001 & 0.000\\
resnet50\_PGD\_ns\_8\_ss\_3 & 0.000 & 0.000 & 0.000 & 0.000 & 0.733 & 0.027 & 0.006 & 0.000 & 0.000 & 0.000 & 0.199 & 0.028 & 0.007\\
resnet50\_PGD\_ns\_16\_ss\_1 & 0.000 & 0.000 & 0.000 & 0.000 & 0.000 & 0.560 & 0.002 & 0.000 & 0.000 & 0.000 & 0.000 & 0.424 & 0.013\\
resnet50\_PGD\_ns\_16\_ss\_3 & 0.000 & 0.000 & 0.000 & 0.000 & 0.004 & 0.081 & 0.297 & 0.000 & 0.000 & 0.000 & 0.000 & 0.204 & 0.414\\
vgg16\_FGSM\_ss\_1 & 0.073 & 0.207 & 0.005 & 0.006 & 0.000 & 0.000 & 0.000 & 0.706 & 0.001 & 0.002 & 0.000 & 0.000 & 0.000\\
vgg16\_FGSM\_ss\_3 & 0.006 & 0.009 & 0.098 & 0.010 & 0.000 & 0.000 & 0.000 & 0.005 & 0.869 & 0.002 & 0.000 & 0.001 & 0.000\\
vgg16\_PGD\_ns\_8\_ss\_1 & 0.000 & 0.000 & 0.000 & 0.301 & 0.000 & 0.000 & 0.000 & 0.000 & 0.000 & 0.697 & 0.000 & 0.003 & 0.000\\
vgg16\_PGD\_ns\_8\_ss\_3 & 0.000 & 0.000 & 0.000 & 0.000 & 0.154 & 0.019 & 0.004 & 0.000 & 0.000 & 0.000 & 0.752 & 0.060 & 0.012\\
vgg16\_PGD\_ns\_16\_ss\_1 & 0.000 & 0.000 & 0.000 & 0.000 & 0.000 & 0.039 & 0.000 & 0.000 & 0.000 & 0.000 & 0.000 & 0.956 & 0.005\\
vgg16\_PGD\_ns\_16\_ss\_3 & 0.000 & 0.000 & 0.000 & 0.000 & 0.002 & 0.082 & 0.146 & 0.000 & 0.000 & 0.000 & 0.004 & 0.214 & 0.552 \\
\hline
\end{tabular}

    \vspace{6pt}
    \caption{Confusion matrix for Laplace method on test dataset. Column labels are in the same order as row labels. Rows indicate ground truth, columns indicate predicted.}
    \label{tab:conf_lap}
\end{table*}

\begin{table*}[]
    \centering
    \scriptsize

\begin{tabular}{|c|c|c|c|c|c|c|c|c|c|c|c|c|c|c|}
\hline
original\_resized & 0.941 & 0.042 & 0.002 & 0.001 & 0.000 & 0.000 & 0.000 & 0.010 & 0.001 & 0.002 & 0.000 & 0.000 & 0.000\\
resnet50\_FGSM\_ss\_1 & 0.008 & 0.880 & 0.003 & 0.001 & 0.000 & 0.000 & 0.000 & 0.108 & 0.000 & 0.001 & 0.000 & 0.000 & 0.000\\
resnet50\_FGSM\_ss\_3 & 0.001 & 0.001 & 0.726 & 0.001 & 0.000 & 0.000 & 0.000 & 0.000 & 0.269 & 0.002 & 0.000 & 0.000 & 0.000\\
resnet50\_PGD\_ns\_8\_ss\_1 & 0.000 & 0.000 & 0.000 & 0.853 & 0.000 & 0.000 & 0.000 & 0.000 & 0.000 & 0.146 & 0.000 & 0.000 & 0.000\\
resnet50\_PGD\_ns\_8\_ss\_3 & 0.000 & 0.000 & 0.000 & 0.000 & 0.934 & 0.000 & 0.000 & 0.000 & 0.000 & 0.000 & 0.065 & 0.000 & 0.000\\
resnet50\_PGD\_ns\_16\_ss\_1 & 0.000 & 0.000 & 0.000 & 0.000 & 0.000 & 0.958 & 0.000 & 0.000 & 0.000 & 0.000 & 0.000 & 0.042 & 0.000\\
resnet50\_PGD\_ns\_16\_ss\_3 & 0.000 & 0.000 & 0.000 & 0.001 & 0.004 & 0.000 & 0.983 & 0.000 & 0.000 & 0.000 & 0.000 & 0.000 & 0.013\\
vgg16\_FGSM\_ss\_1 & 0.012 & 0.746 & 0.002 & 0.000 & 0.000 & 0.000 & 0.000 & 0.238 & 0.001 & 0.001 & 0.000 & 0.000 & 0.000\\
vgg16\_FGSM\_ss\_3 & 0.001 & 0.000 & 0.379 & 0.001 & 0.000 & 0.000 & 0.000 & 0.001 & 0.617 & 0.002 & 0.000 & 0.000 & 0.000\\
vgg16\_PGD\_ns\_8\_ss\_1 & 0.001 & 0.000 & 0.000 & 0.130 & 0.000 & 0.000 & 0.000 & 0.000 & 0.000 & 0.868 & 0.000 & 0.001 & 0.000\\
vgg16\_PGD\_ns\_8\_ss\_3 & 0.000 & 0.000 & 0.000 & 0.000 & 0.091 & 0.000 & 0.000 & 0.000 & 0.000 & 0.000 & 0.909 & 0.000 & 0.000\\
vgg16\_PGD\_ns\_16\_ss\_1 & 0.000 & 0.000 & 0.000 & 0.000 & 0.000 & 0.053 & 0.000 & 0.000 & 0.000 & 0.000 & 0.000 & 0.947 & 0.000\\
vgg16\_PGD\_ns\_16\_ss\_3 & 0.000 & 0.000 & 0.000 & 0.000 & 0.000 & 0.000 & 0.002 & 0.000 & 0.000 & 0.000 & 0.000 & 0.000 & 0.998 \\
\hline
\end{tabular}

    \vspace{6pt}
    \caption{Confusion matrix for co-occurrence method on test dataset. Column labels are in the same order as row labels. Rows indicate ground truth, columns indicate predicted.}
    \label{tab:conf_co_occur}
\end{table*}

\subsection{Discussion}

Across all tasks, the co-occurrence preprocessing tended to perform the best. Especially noteworthy is the difference in performance of the direct method between the classification tasks and the parameter estimation tasks. Of the three, the direct provides the most information too the neural network. While this led to good results on the training classes, the information bottleneck provided by the co-occurrence and Laplace functions may help reduce overfitting.

\section{Conclusion and Future Directions}

In this work, we presented several methods for attribution and parameter estimation of select adversarial attacks. This combination of detection, attribution, and parameter estimation was accomplished using a single pass through a multi-class neural network, trained on a sampling of several common adversarial attacks.

While our model was demonstrated effective against several attacks not seen in the training sets, there are a variety of additional attacks to be considered. Furthermore, our method of parameter interpolation is inherently limited to estimating values only within the range of values in the training set. Creation of a more robust model for real-world deployment would require a broader sampling of attack methods, target networks, and attack parameters.

\begin{table}[]
    \centering
    \begin{tabular}{|c|c|c|c|}
        \hline
        Meta Classes & Direct & Laplace & Co-occur \\
        \hline
        Binary Detection & 0.921 & 0.955 & \textbf{0.970} \\
        Full attribution & \textbf{0.907} & 0.834 & 0.808 \\
        Original, ResNet, VGG & \textbf{0.925} & 0.834 & 0.865 \\
        Original, FGSM, PGD & 0.919 & 0.961 & \textbf{0.978} \\
        Full Classification & \textbf{0.928} & 0.766 & 0.835 \\
        \hline
    \end{tabular}
    \vspace{6pt}
    \caption{Mean average precision on different meta classification tasks. Full Attribution denotes classification between the original, FGSM ResNet, FGSM VGG, PGD ResNet, and PGD classes. Full Classification refers to accuracy across all 13 classes in the training set. }
    \label{tab:classification}
\end{table}

\begin{table}[]
    \centering
    \begin{tabular}{|c|c|c|c|}
    \hline
         & Direct & Laplace & Co-Occur \\
        \hline
        FGSM step size & 0.491 & \textbf{0.469} & 0.509 \\
        PGD step size & 0.567 & 0.680 & \textbf{0.535} \\
        PGD number of steps & 4.17 & 3.66 & \textbf{3.42} \\
        \hline
    \end{tabular}
    \vspace{6pt}
    \caption{Root Mean Squared Error (RMSE) for parameter estimation. Step sizes are sampled from \{1,2,3\}, and number of steps sampled from \{8,12,16\}.}
    \label{tab:estimation}
\end{table}

%%%%%%%%%%%%%%%%%%%%%%%%%%%%%%%%%%
% Bibliography
%%%%%%%%%%%%%%%%%%%%%%%%%%%%%%%%%%

{\small
    \bibliographystyle{ieee}
    \bibliography{red}
}

%%%%%%%%%%%%%%%%%%%%%%%%%%%%%%%%%%
% Biography
%%%%%%%%%%%%%%%%%%%%%%%%%%%%%%%%%%

\section{Acknowledgements}

This material is based upon work supported by the Defense Advanced Research Projects Agency (DARPA) under Agreement No. HR00111990080. The views, opinions and/or findings expressed are those of the author and should not be interpreted as representing the official views or policies of the Department of Defense or the U.S. Government.

\begin{biography}

\textbf{Michael Goebel} received his B.S. and M.S. degrees in Electrical Engineering from Binghamton University in 2016 and 2017. He is currently a PhD student in Electrical Engineering at University of California Santa Barbara.

\textbf{Jason Bunk} received his B.S. degree Computational Physics
from the University of California, San Diego in 2015, and his M.S.
degree in Electrical and Computer Engineering from the University of California, Santa Barbara in 2016. He is currently a Research Staff Member at Mayachitra Inc., Santa Barbara, CA. His recent research efforts include adversarial attacks and defenses, applying deep learning techniques to media forensics, and active learning with neural networks for video activity detection.

\textbf{Srinjoy Chattopadhyay} received the B.Tech. degree in electronics and electrical communication engineering and the M.Tech. degree in telecommunication systems engineering from IIT Kharagpur, India, in 2013, and the Ph.D. degree from the Department of Electrical and Computer Engineering, North Carolina State University, USA in 2020.  He is currently a Research Staff Member at Mayachitra Inc., Santa Barbara, CA. His current research interests include wireless communications, computer networks, spectral theory of
graphs, and network design problems on multilayer interdependent networks

\textbf{Lakshmanan Nataraj} received his B.E degree from Sri Venkateswara College of Engineering, Anna university in 2007, and the Ph.D. degree in the Electrical and Computer Engineering from the University of California, Santa Barbara in 2015. 
He is currently a Senior Research Staff Member at Mayachitra Inc., Santa Barbara, CA. 
His research interests include malware analysis and image forensics. 

\textbf{Shivkumar Chandrasekaran} received his Ph.D. degree in Computer Science from Yale
University, New Haven, CT, in 1994. He is a Professor in the Electrical and Computer
Engineering Department, University of California, Santa Barbara. His research interests are in Computational Mathematics

\textbf{B. S. Manjunath} received the Ph.D. degree in Electrical Engineering from the University of Southern California in 1991. He is currently a Distinguished Professor at the ECE Department at the University of California at Santa Barbara. He has co-authored about 300 peer-reviewed articles. His current research interests include image processing, computer vision and biomedical image analysis.

\end{biography}

\end{document}